\documentclass[11pt]{article}
\usepackage{authblk}
\usepackage[a4paper, margin=1in]{geometry}
\usepackage{mathpazo}
\usepackage[semibold]{sourcesanspro}
\usepackage{titlesec}
\titleformat*{\section}{\Large\sffamily}
\titleformat*{\subsection}{\large\sffamily}
\titleformat*{\subsubsection}{\sffamily}
\usepackage{times}
\usepackage{bm}
\usepackage{textcomp}
\usepackage{tikz}
\usetikzlibrary{quantikz2, decorations.pathreplacing}
\usepackage{graphicx}
\usepackage{bm}
\allowdisplaybreaks 
\usepackage{braket}
\usepackage{physics}
\usepackage{subdepth}
\usepackage[breaklinks]{hyperref}
\hypersetup{colorlinks=true, linkcolor=blue, citecolor=blue, filecolor=blue, urlcolor=blue}
\usepackage{algorithm2e}
\usepackage{mathtools}
\usepackage{amsthm}
\newtheorem{theorem}{Theorem}

\theoremstyle{definition}

\newcommand{\RN}[1]{\textup{\uppercase\expandafter{\romannumeral#1}}}

\def\numx#1e#2{{#1}\mathrm{e}{#2}}
\begin{document}

\RestyleAlgo{ruled}

\SetKwComment{Comment}{/* }{ */}
\SetKwInput{Input}{Input}
\SetKwInput{Output}{Output}
\SetKwInput{Return}{Return}
\SetKwInput{Guarantee}{Guarantee}
\SetKwFor{For}{for}{do}{endfor}
\DontPrintSemicolon

\title{The table maker's quantum search}
\author{Benjamin C. A. Morrison}
\affil{Institut quantique \& D\'{e}partement de physique \\ Universit\'{e} de Sherbrooke, Sherbrooke, J1K 2R1, Canada}

\author{Stefanos Kourtis}
\affil{Institut quantique \& D\'{e}partement de physique \& D\'{e}partement d'informatique \\ Universit\'{e} de Sherbrooke, Sherbrooke, J1K 2R1, Canada}
\affil{Mila -- Quebec AI Institute \& IVADO - Institut de valorisation des données \\ Montréal, H2S 3H1, Canada}

\maketitle

\begin{abstract}
We show that quantum search can be used to compute the hardness to round an elementary function, that is, to determine the minimum working precision required to compute the values of an elementary function correctly rounded to a target precision of $n$ digits for all possible precision-$n$ floating-point inputs in a given interval. For elementary functions $f$ related to the exponential function, quantum search takes time $\tilde O(2^{n/2} \log (1/\delta))$ to return, with probability $1-\delta$, the hardness to round $f$ over all $n$-bit floating-point inputs in a given binade. For periodic elementary functions in large binades, standalone quantum search yields an asymptotic speedup over the best known classical algorithms and heuristics. We then estimate the resources required for a fault-tolerant implementation of the proposed algorithm for the $\sin$ and $\cos$ functions in double precision. We find that, although the algorithm can in principle compete with the fastest known practical method for computing the hardness to round over all binades in the format, it requires qubit coherence times that are unrealistically long for present technology.
\end{abstract}

\section{Introduction}
\label{sec:introduction}

Quantum search algorithms based on Grover iteration~\cite{grover1996, Brassard_2002} promise to accelerate the solution of combinatorial search and optimization problems. When Grover's algorithm and its variants are used as standalone solvers, they generically achieve a quadratic speedup over exhaustive classical search. For example, hard Boolean satisfiability instances on $n$ variables can be solved in time $\tilde O(2^{n/2})$, as opposed to $O(2^n)$ for exhaustive classical search. On the other hand, decades of research has yielded powerful classical algorithms that beat the scaling of standalone Grover search for most problems of practical interest by exploiting the problem structure. To obtain a quantum advantage for these problems, Grover iteration must be used as a subroutine inside suitable classical algorithms~\cite{Brassard_2002, ambainis2004, furer2008}.

Are there problems of practical interest for which standalone Grover search beats the best known classical algorithm? For such a quantum advantage to occur, the search problem must be (approximately) devoid of structure that can be exploited by a classical algorithm. Relatively few such problems are known. One is symmetric-key cryptography~\cite{Bernstein2009, grassl2016}. However, simply increasing the key length is sufficient to thwart attacks by standalone quantum search, due to its superpolynomial scaling. Another essentially structure-free problem is circuit satisfiability (CSAT)~\cite{Arora_Barak_2009}. No known classical algorithm is meaningfully faster than exhaustive search in the worst case. In practice, industrial CSAT instances do have structure and it is found empirically that they can be solved with methods much faster than exhaustive search by mapping to equilogical Boolean satisfiability formulas~\cite{Amla2005, Vizel2015}.

Here we study the problem of determining the working precision required to compute the correctly rounded value of a function at finite-precision floating-point inputs. Correct rounding is of practical importance in computing. Compounded rounding errors can give rise to large deviations between the computed and exact values of a quantity. Examples of adverse consequences of incorrect rounding include wildly erroneous index values of the Vancouver stock exchange in the early 1980s~\cite{mccullough1999, nievergelt2000rounding} and the failure of a US Army Patriot array to intercept a Scud missile during the Gulf War~\cite{or1992patriot, schmitt1991after}. These mishaps highlight the importance of verifying that computer programs round correctly.

For transcendental functions, such verification is in general difficult: it is not known to which working precision $p>n$ we have to evaluate the function output for a given precision-$n$ floating-point input before we can guarantee correct rounding to precision $n$~\cite{muller2009solving, brisebarre2025}. This problem is known as the table maker's dilemma~\cite{kahan2004}. The working precision required to guarantee correctly rounded evaluation of a function $f$ for any precision-$n$ floating-point input $x$ in an interval $I$ is called the hardness to round $f$ in $I$, denoted $\mathrm{htr}_{f,I}(n)$. The difficulty in determining $\mathrm{htr}_{f,I}(n)$ arises from bad rounding cases, inputs for which the function, computed to working precision $p>n$, evaluates to a number that is too close to a decision boundary, so that we cannot unequivocally decide whether we should round to one or the other of two consecutive precision-$n$ floating-point numbers. Although a $O(n^2)$ upper bound is known for the hardness to round the exponential and related functions, such as the logarithm and trigonometric functions~\cite{nesterenko2000approximationvaluesexponentialfunction, brisebarre2025}, the hidden prefactors are enormous and the bound is not useful in practice. With the non-rigorous assumption that the digits of the value of a function are uniformly distributed random numbers, we can estimate that the hardness to round to $n$ digits is roughly $2n$~\cite{muller2009solving}. Although this estimate matches empirical evidence well, it is not a worst-case bound.

No known polynomial-time algorithm computes the hardness to round a transcendental function exactly. The first nontrivial algorithm for computing the hardness to round a function $f$ over precision-$n$ binary floating-point inputs in a given binade scales as $\tilde O(2^{2n/3})$~\cite{Lefevre2005}. This algorithm assumes a sufficient number of consecutive floating-point numbers in the binade to approximate $f$ by a polynomial of degree 1. By increasing the approximating polynomial degree, the performance can be improved to $\tilde O(2^{n/2})$, at the expense of rendering the algorithm incomplete, i.e., a heuristic~\cite{Stehle2006, slz2005}. The reason is that the algorithm of~\cite{Stehle2006, slz2005} uses Coppersmith's method to find small roots of bivariate polynomials~\cite{Coppersmith1996}, a task for which the method is only known to be heuristic.

Both the aforementioned algorithms assume that a low-degree polynomial approximation is viable. For periodic elementary functions, this assumption does not always hold: consecutive floating-point numbers in large binades may even belong to different periods of the function. Ref.~\cite{Hanrot2007} overcomes this issue via a preprocessing step that uses the periodicity of the function to reorder inputs into an arithmetic progression. Then, the algorithm of~\cite{Lefevre2005} or the heuristics of~\cite{Stehle2006, slz2005} can be applied, leading to runtimes $\tilde O(2^{4n/5})$ and $\tilde O(2^{(7-2\sqrt{10})n}) \approx \tilde O(2^{0.676n})$, respectively.

\section{Contribution}
\label{sec:contribution}

The scalings of the best known classical algorithms and heuristics for computing the hardness to round cited above indicate that this task may be a good application of quantum search. The following theorem and algorithm formalize this intuition.

\begin{theorem} \label{theorem:htr}
    For any elementary function $f$ evaluated at precision-$n$ binary floating-point inputs $x\in I$ where $I=[2^e,2^{e+1})$, Algorithm~\ref{alg:qhtr} returns $l = \min(\{ \mathrm{htr}_{f,I}(n), p_{\max} \})$ with probability $1-\delta$ by making $O(2^{n/2} \log\frac{\lfloor \log p_{\max} \rfloor + 1}{\delta})$ calls to the oracles $\mathcal{O}^{f,\bullet}_{n,e}$. Furthermore, if $p_{\max} = O(\mathrm{poly}(n))$ is an upper bound for $\mathrm{htr}_{f,I}(n)$, Algorithm~\ref{alg:qhtr} returns $\mathrm{htr}_{f,I}(n)$ with probability $1-\delta$ in time $\tilde O(2^{n/2} \log (1/\delta))$.
\end{theorem}

In Algorithm~\ref{alg:qhtr}, the Grover operator $\mathcal{O}^{f,p}_{n,e}$ implements an arithmetic circuit that evaluates $f(x)$ up to $p$ significant digits, coherently for all precision-$n$ binary floating-point inputs $x$. A membership oracle then marks inputs $x$ that yield bad rounding cases for $f(x)$ at precision $n$. We define $\mathcal{O}^{f,\bullet}_{n,e} = \{ \mathcal{O}^{f,p}_{n,e} \}_{p=n+1}^{p_{\max}}$. If $S$ is the set of bad rounding cases of a function $f$, we are searching for the minimum working precision $p$ for which $S$ is empty. Therefore, instead of searching for elements of a nonempty $S$ -- the typical use case for Grover search -- we are instead asking a question about $|S|$. Whenever $S$ is nonempty, measurement of the output state returns bad rounding cases.

\begin{algorithm}[t]
\caption{Quantum search for $\mathrm{htr}_{f,I}(n)$, $I=[2^e,2^{e+1})$}\label{alg:qhtr}
\Input{Precision $n$, oracle circuits $\mathcal{O}^{f,\bullet}_{n,e}$, upper bound $p_{\max}$, failure probability $\delta$}
\Output{$\min(\{ \mathrm{htr}_f(n), p_{\max} \})$}
$l \gets n+1$ \;
$r \gets p_{\max} $ \;
$\delta' \gets \delta / (\lfloor \log p_{\max} \rfloor + 1)$ \;
\While{ $l<r$ }{
$p \gets l + \frac{\lfloor r-l \rfloor}{2}$ \;
$k \gets \mathrm{QSearch}(n,\mathcal{O}^{f,p}_{n,e}, \delta')$ \;
\eIf{$k>0$}{
$l \gets p+1$
}{
$r \gets p$
}
}
\Return{$l$}
\end{algorithm}
\phantom{a}

The subroutine QSearch in Algorithm~\ref{alg:qhtr} is the quantum search algorithm~\cite{grover1996, Boyer_1998}. Given an oracle $\mathcal{O}_S$ that marks the elements of a set $S \subseteq \{ 0,1 \}^n$ and $0< \delta' < 1$, $\mathrm{QSearch}(n,\mathcal{O}_s, \delta')$ returns, with probability $1-\delta'$, 1 whenever $|S|>0$ and 0 whenever $|S|=0$. Algorithm~\ref{alg:qhtr} is then essentially a binary search over the working precision $p \in [n+1,p_{\max}]$ for the smallest integer $p^\star$ such that no marked item is found. If such a $p^\star$ exists in the searched interval, it is the hardness to round $f$. This approach works because the number of bad rounding cases is a non-increasing function of $p$. Finally, to achieve an overall failure rate $\delta$, we need the failure rate of each of the $\lfloor \log p_{\max} \rfloor + 1$ runs of QSearch to scale as $\delta ' = \delta / (\lfloor \log p_{\max} \rfloor + 1)$, thus obtaining the scaling cited in Theorem~\ref{theorem:htr}.

For functions related to the exponential function, such as the logarithm and trigonometric functions, a theorem of Nesterenko and Waldschmidt implies that we can take $p_{\max} = Cn^2$ for a sufficiently large and straightforwardly computable constant $C$~\cite{nesterenko2000approximationvaluesexponentialfunction, brisebarre2025}. Therefore, for these functions, Algorithm~\ref{alg:qhtr} works in time $\tilde O(2^{n/2} \log (1/\delta))$ in any binade. We thus obtain a modest asymptotic speedup over the best known classical algorithms and heuristics for computing the hardness to round periodic elementary functions in large binades~\cite{Hanrot2007}. Moreover, by initializing the $d$-bit exponent of inputs $x$ in a superposition of all possible values for a precision-$n$ floating-point format and suitably modifying the Grover operator, we can compute the hardness to round a function over all possible $2^{n+d}$ floating-point inputs in the format in time $\tilde O(2^{(n+d)/2} \log (1/\delta))$.

That standalone quantum search asymptotically outperforms the best known classical methods for a problem of some practical importance is interesting in its own right as a theoretical result. On the other hand, whether quantum search can be advantageous in practice for this task depends on the precise scaling of the number of quantum operations required for a fault-tolerant implementation of Algorithm~\ref{alg:qhtr}. In Section~\ref{sec:resources}, we estimate the resources needed to apply Algorithm~\ref{alg:qhtr} to the $\sin$ and $\cos$ functions in double precision. We find that, although the algorithm can in principle compete with the fastest known practical method for computing the hardness to round over all binades in the format~\cite{lefevre2026}, it requires qubit coherence windows that are unrealistically long for present technology.

\section{Preliminaries and notation}
\label{sec:preliminaries}

Let $\mathbb{Z}, \mathbb{Q}, \mathbb{R}$, and $\mathbb{C}$ denote the sets of integer, rational, real, and complex numbers respectively. A number $x$ is rational if $x = p/q$, with $p, q \in \mathbb{Z}$ and $q\not=0$. A number $z \in \mathbb{C}$ is algebraic if there exists a nonzero polynomial $P$ with integer coefficients such that $P(z)=0$. A number $z \in \mathbb{C}$ is transcendental if it is not algebraic. Arithmetic operations are the operations of addition, subtraction, multiplication, and division.

In this work we will be dealing exclusively with univariate functions. A complex-valued function $f$ is an algebraic function if there exists a bivariate polynomial $P$ with integer coefficients such that for all $z$ in the domain of $f$ we have $P(z,f(z)) = 0$. A function is transcendental if it is not algebraic. A function is elementary if it can be defined by applying finitely many arithmetic operations and function compositions to polynomial, exponential, logarithm, and constant functions. Elementary functions have derivatives of all orders that are algorithmically computable. For simplicity, below we consider exclusively real numbers and real-valued functions, but our results generalize to the complex plane. 

A binary floating-point number $x$ is a triple $(s,m,e)$ such that
\begin{equation}
    x = (-1)^s \cdot m \cdot 2^e \,, \label{eq:fp}
\end{equation}
where $s\in \{ 0,1 \}$ is the sign bit, $m \geq 0$ is the significand or mantissa, and $e \in \mathbb{Z}$ is the exponent. The precision of $x$ is the number of variable significant digits of $m$. The possible values that a precision-$n$ floating-point number can take depend on the semantics chosen for representing $m$. The binary floating-point formats of the IEEE 754 standard~\cite{ieee754} define the significand of normal numbers as
\begin{equation}
    m = 1.m_1 m_2 \dots m_{n} = \sum_{j=0}^{n} m_j 2^{-j}  \,, \label{eq:mantissa}
\end{equation}
that is, we fix the implicit integer bit $m_0 = 1$ and use $n$ bits to define the fractional part of $m$, called the fraction, so that $m \in [1,2)$. In the following, when we refer to ``the $n$-bit significand of $x$'', we mean that $m(x)$ has $n$ variable digits after the implicit one. For fixed $e$ and $s=0$, there are $2^{n}$ precision-$n$ binary floating-point numbers in the binade $[2^e,2^{e+1})$. If $e$ is also variable and defined as a $d$-bit integer, $x$ can take $2^{n+d+1}$ values\footnote{In the IEEE 754 standard, some of these values are reserved for special purposes: the values $e=m=0$ encode the signed zeros $\pm0$; $e=0, m\not=0$ the subnormal numbers; $e=2^d-1, m=0$ the positive and negative infinity $\pm\infty$; and $e=2^d-1, m\not=0$ the Not-a-Number value NaN.}. For convenience, we define the functions $s(x)$, $m(x)$, and $e(x)$ that return the sign bit, significand, and exponent, respectively, of a floating-point number $x$.

From Eqs.~\eqref{eq:fp} and~\eqref{eq:mantissa} we see that binary floating-point numbers are rational numbers whose denominator is a power of 2. For rational numbers whose denominator is not a power of 2, the expansion~\eqref{eq:mantissa} may not terminate for any finite $n$ (take, for example, the number 4/3). The same holds for irrational numbers. In floating-point arithmetic, for any number $x$ whose binary floating-point representation does not terminate, we first compute sufficiently many significand digits to obtain an approximation $\tilde x$ and then round $\tilde x$ to the target precision, that is, we replace the result with a nearby floating-point number. Which nearby floating-point number we pick depends on the rounding mode chosen. Common rounding modes are rounding up/down, which returns the closest floating-point number greater/smaller than the input, and rounding to nearest, which returns the floating-point number closest to the input. In what follows, we discuss only the round-to-nearest mode for illustration purposes, but all our results translate straightforwardly to other rounding modes.

We now describe the ``round to nearest, ties to even'' rounding mode, which is the most commonly used in binary floating-point arithmetic. For a precision-$p$ floating-point number $x$, let $\circ_n(x)$ denote its value rounded to the nearest precision-$n$ floating-point number, with $n<p$. We call $p$ the working precision and $n$ the target precision. To compute $\circ_n(x)$, we examine the $p-n+1$ least significant digits of the significand $m = m(x)$, starting with the guard digit $m_{n+1}$:
\begin{enumerate}
    \item If $m_{n+1} = 0$, then $\circ_n(x) = (-1)^{s(x)} \cdot 1.m_1 \dots m_n \cdot 2^e$, that is, we simply truncate the $p-n$ least significant digits of the significand (i.e., round toward zero).
    \item If $m_{n+1} = 1$, then:
    \begin{enumerate}
        \item If $m_j = 1$ for any $j>n+1$, then  $\circ_n(x) = (-1)^{s(x)} \cdot (1.m_1 \dots m_n + 2^{-n}) \cdot 2^e$ (i.e., round away from zero).
        \item If $m_j = 0$ for all $j>n+1$, then $x$ is exactly halfway between two precision-$n$ floating-point numbers and we need to break the tie somehow. The ``ties to even'' rule is:
        \begin{enumerate}
            \item If $m_n = 0$, we round toward zero: $\circ_n(x) = (-1)^{s(x)} \cdot 1.m_1 \dots m_n \cdot 2^e$.
            \item If $m_n = 1$, we round away from zero: $\circ_n(x) = (-1)^{s(x)} \cdot (1.m_1 \dots m_n + 2^{-n}) \cdot 2^e$.
        \end{enumerate}
    \end{enumerate}
\end{enumerate}
Note that if $p=n+1$, then $m_j = 0$ implicitly for all $j>n+1$.

As mentioned previously, for numbers $x$ whose binary floating-point representation does not terminate, we instead use a computation method that returns an approximation $\tilde x$ with working precision $p$, so that $\tilde x$ is guaranteed to be within $2^{-p}$ of $x$. We say that this method implements correct rounding if $\circ_n(\tilde x) = \circ_n(x)$.

The same reasoning holds for transcendental functions $f$, since the exact value $f(x) \in \mathbb{R}$ is not representable as a finite-precision floating point number in general. In floating-point arithmetic, we instead evaluate a function $\tilde f$ whose value $\tilde f(x)$ is a precision-$p$ floating-point number within $2^{e(f(x))-p}$ of $f(x)$, where $p > n$. This is done by summing a convergent series for $f$, such as a Taylor or arithmetic-geometric mean series~\cite{Brent_Zimmermann_2010}. A method that evaluates $f$ via $\tilde f$ implements correct rounding to precision $n$ if $\circ_n(f(x)) = \circ_n(\tilde f(x))$ for all precision-$n$ floating-point inputs $x$.

It is not known a priori to what precision we have to evaluate $\tilde f$ to ascertain correct rounding to precision $n$~\cite{muller2009solving}. Consider the following example. Suppose that we compute the precision-$p$ floating-point number $\tilde f(x)$, with $\tilde m = m(\tilde f(x))$ and $e(\tilde f(x))=e(f(x))=0$, to obtain
\begin{equation}
    \tilde m = 1. \tilde m_1 \tilde m_2 \dots \tilde m_{n-1} 1 1 0 \dots 0 \,.
\end{equation}
We see that the digit $\tilde m_n = 1$ is followed by the guard digit $\tilde m_{n+1} = 1$, followed by $p-n-1$ zeros. For $\tilde f(x)$ to be correctly rounded to precision $n$, we must ascertain that $\tilde m_j = m_j$ for $j=0,\dots,n$, where $m=m(f(x))$ is the infinitely precise significand of $f(x)$. This is not necessarily the case: $\tilde m$ and $\tilde m' = 1. \tilde m_1 \tilde m_2 \dots \tilde m_{n-1} 1 0 1 \dots 1$ define floating-point numbers that are both within $2^{-p}$ of $f(x)$. When rounding to nearest with ties to even, $\tilde m_n$ would be incremented to obtain $\tilde m_n=0$, while $\tilde m_n'$ would be left as $\tilde m_n'=1$. Therefore, we cannot decide whether $\tilde m_n = m_n$, which, in turn, means that we cannot ascertain whether $\circ_n(f(x)) = \circ_n(\tilde f(x))$. Now, suppose we compute $\tilde f(x)$ up to higher precision of $p+1$ bits, that is, we approximate $f(x)$ to within $2^{-(p+1)}$ by summing more terms in the series expansion of $f$. If we obtain
\begin{equation}
    \tilde m = 1. \tilde m_1 \tilde m_2 \dots \tilde m_{n-1} 1 1 0 \dots 01 \,,
\end{equation}
then the only precision-$(p+1)$ floating-point numbers within $2^{-(p+1)}$ of $f(x)$ have significands that agree with the infinitely precise significand of $f(x)$ on up to $n$ significant digits. Therefore, we can ascertain that $\circ_n(f(x)) = \circ_n(\tilde f(x))$.

For bad rounding cases $x$, the last $p-n-1$ digits of $m(\tilde f(x))$ evaluate to $00\dots0$ or $11\dots1$. Such long runs of 0s or 1s can arise in two cases. The first (trivial) case arises when $f(x)$ evaluates exactly to a floating-point number of precision smaller than $p$, such as the function $\log_2$ evaluated at integer powers of 2. For most common transcendental functions, these exceptional inputs are well-known and can be efficiently specified and excluded~\cite{muller2009solving}. The second case is the one illustrated in the example above, when there are more than one precision-$p$ floating-point numbers within $2^{-p}$ of $f(x)$. A concrete example is the single-precision (binary32 format of IEEE 754) input
\begin{equation}
    x = 1.00111011101100100011010_2 \cdot 2^{-1}
\end{equation}
with 24-bit significand, for which the function $f= 2 \sin$ evaluates to
\begin{align}
    2\sin(x) = 1.&000110111101000110110010 \nonumber\\ 
    &111111111111111111111000\dots_2 \,,
\end{align}
that is, the 24th digit of $m(f(x))$ is 1, followed by the guard digit 0, followed by 21 consecutive 1s. Therefore, to guarantee correct rounding for this input, we have to approximate $f(x)$ up to a precision of at least $24+21=45$ bits.

Given a function $f$ and target precision $n$, up to what precision $p$ do we need to evaluate $\tilde f(x)$ in order to be sure that $\circ_n(f(x)) = \circ_n(\tilde f(x))$? This question is known as the table maker's dilemma~\cite{kahan2004}. The smallest $p$ for which $\circ_n(f(x)) = \circ_n(\tilde f(x))$ for all precision-$n$ floating-point inputs $x \in I$, where $I$ is some interval, is called the hardness to round $f$ in $I$ and denoted $\mathrm{htr}_{f,I}(n)$. We denote the hardness to round $f$ over all precision-$n$ floating-point inputs $\mathrm{htr}_{f}(n)$.

\section{Grover search and membership oracle}
\label{sec:oracle}

For completeness, we give an overview of Grover search~\cite{grover1996} (see~\cite{Nielsen_Chuang_2010} for a complete introduction). Given a search problem over $n$ Boolean variables, the Grover algorithm in its simplest form is the repeated application of the unitary operator $\mathcal{G} = \mathcal{D} \mathcal{O}$ to the $n$-qubit quantum state $\ket{+}^{\otimes n} = 2^{-n/2} \sum_{k=0}^{2^n-1} \ket{k}$, where $k$ runs over all $n$-bit strings. The membership oracle $\mathcal{O}$ acts as $\mathcal{O} \ket{k} = - \ket{k}$ when $k$ is a solution and as $\mathcal{O} \ket{k} = \ket{k}$ otherwise. We describe the circuit construction of a membership oracle for bad rounding cases below. The diffusion operator $\mathcal{D} = 2 (\ket{+}\bra{+})^{\otimes n} - I$, where $I$ is the identity operator, is constructed with $O(n)$ Hadamard and classical controlled gates. After $r$ iterations, the algorithm produces the state
\begin{equation}
    \mathcal{G}^r \ket{+}^{\otimes n} = \cos \left( \frac{2r+1}{2} \theta \right) \ket{\bar S} + \sin \left( \frac{2r+1}{2} \theta \right) \ket{S} \,,
\end{equation}
where $\ket S = |S|^{-1/2} \sum_{k \in S} \ket{k}$ is the equal superposition of all states in the solution set $S$, $\ket{\bar S} = |2^n - S|^{-1/2} \sum_{k \notin S} \ket{k}$ the equal superposition of non-solutions, and $\cos\theta/2 = \sqrt{(2^n - |S|)/2^n}$. Measurement of the resulting state yields a solution to the search problem with probability $\sin^2 \left( \frac{2r+1}{2} \theta \right)$. When $\mathcal{O}$ can be expressed as a poly-sized quantum circuit, the worst-case runtime for either returning a solution, if there is one, or asserting that there are no solutions with high probability is $\tilde O(2^{n/2})$~\cite{Boyer_1998}.

The membership oracles for Algorithm~\ref{alg:qhtr} are composed of two circuits. The first one is a reversible arithmetic circuit that, on input $x$ a precision-$n$ floating-point number with exponent $e$, computes a precision-$p$ approximation $\tilde f(x)$ to $f(x)$, correct up to additive error $2^{-p}$, with $p>n$. This is a standard task in arbitrary-precision arithmetic and can be implemented with multiple methods in time polynomial in $p$~\cite{Brent_Zimmermann_2010}, with the (asymptotically) fastest known being the arithmetic-geometric mean method that runs in time $O(M(p)\log p)$, where $M(p)$ is the cost of multiplying two $n$-bit integers. Turning any of these methods into a reversible circuit incurs at most polynomial overhead in time and ancilla bits~\cite{bennett1989, levine1990}. After computation, any ancilla bits used are returned to their initial state by uncomputation. Here we are interested in comparisons with classical methods whose scaling is typically evaluated in a fixed binade. We therefore choose to hard-code the input exponent $e$ into the arithmetic circuits $C^{f,p}_{n,e}$ and give only the $n$-bit significand $m(x)$ as input. Similarly, although it implicitly evaluates $e(\tilde f(x))$, the circuit outputs only the $n$-bit significand $m(\tilde f(x))$, since that is all we need to detect bad rounding cases. To further lighten the notation, we also omit sign bits and the implicit integer bit of the significands. With this convention, the initial state for the quantum search algorithm is the equal superposition of all $2^n$ precision-$n$ significands, prepared as $\ket{+}^{\otimes n}$.

The second circuit $C^{\mathrm{bad}}_{n,p}$ tests if the last bits of $\tilde f(x)$ form a long string of 0s or 1s, corresponding to a bad rounding case. Specifically, the strings of interest are $110\dots0$ or $101\dots1$ in the case of rounding to nearest with ties to even\footnote{When  rounding down or up, the bad cases terminate in $00\dots0$ or $11\dots1$.}. A circuit for a given rounding mode tests whether the last $p-n+1$ bits of $\tilde f(x)$ are in the corresponding state and, in the affirmative, flags the state $x$ by conditionally flipping a register, which in this case is a single qubit prepared in the state $\ket{-}$. The conditional operation can be implemented with $O(n)$ Toffoli gates and $O(n)$ ancilla qubits and is uncomputed after the state has been flagged, returning all ancillas to their initial state.

The action of the circuits $C^{f,p}_{n,e}$ and $C^{\mathrm{bad}}_{n,p}$ is then
\begin{align}
    C^{f,p}_{n,e} & \ket{m(x)} \ket{0}^{\otimes p} \ket{0}^{\otimes w} = \ket{m(x)} \ket{m(\tilde f(x))} \ket{0}^{\otimes w} \,, \\
    C^{\mathrm{bad}}_{n,p} & \ket{m(x)} \ket{m(\tilde f(x))} \ket{0}^{\otimes w} \ket{-} = \nonumber\\ 
    & = (-1)^{b_n(\tilde f(x))} \ket{m(x)} \ket{m(\tilde f(x))} \ket{0}^{\otimes w} \ket{-} \,,
\end{align}
where $w$ is the number of ancilla bits required by $C^{f,p}_{n,e}$ and $C^{\mathrm{bad}}_{n,p}$, and $b_n(y)$ is an indicator function that evaluates to 1 whenever the floating-point number $y$ is a bad case for rounding to precision $n$.

With the above, we can define the membership oracle for bad rounding cases as
\begin{equation}
    \mathcal{O}^{f,p}_{n,e} = C^{f,p \mathsf{T}}_{n,e} C^{\mathrm{bad}}_{n,p} C^{f,p}_{n,e} \,.
\end{equation}
The action of the oracle is thus
\begin{align}
    \mathcal{O}^{f,p}_{n,e} & \ket{m(x)} \ket{0}^{\otimes p} \ket{0}^{\otimes w} \ket{-} = \nonumber\\
    & = (-1)^{b_n(\tilde f(x))} \ket{m(x)} \ket{0}^{\otimes p} \ket{0}^{\otimes w} \ket{-} \,,
\end{align}
as required for the Grover iteration.

There are some subtleties to address. First, some functions may evaluate exactly to a precision-$n$ floating-point number for some inputs. In such cases, the significand $m(\tilde f(x))$ may terminate in $110\dots0$ (or a string relevant for detecting bad cases for another rounding mode), even though $x$ is not a bad rounding case. Such numbers are exceptional and can be dealt with efficiently. For example, for trigonometric functions $\sin$ and $\cos$, Niven's theorem~\cite{Lehmer1933, Niven1956} implies that the only finite-precision floating-point $x$ for which these functions evaluate to a rational number is $x=0$. Similar results hold for other elementary functions, such as $\ln$ and $\exp$. Some elementary functions, such as power, root, and logarithm functions, evaluate exactly to floating-point numbers with fewer than $n$ significant digits for many inputs (e.g., integers), but these are known and can be efficiently detected and excluded from oracle marking with additional circuit logic~\cite{muller2009solving}. We assume that this logic is included in $C^{f,p}_{n,e}$. Second, we may have $e(x) \not= e(\tilde f(x))$, although both exponents have the same length $d$ and format. An evaluation of $f(x)$ that yields $e(\tilde f(x))$ outside the range representable by a $d$-bit integer leads to an under- or overflow. Such cases can be efficiently detected by exception handling in $C^{f,p}_{n,e}$, as in the IEEE floating-point standard. Because of this, and since underflows and overflows do not constitute bad rounding cases, we ignore them in our discussion. For the same reason, we also ignore inputs corresponding to subnormal numbers, infinities, and NaN.

We conclude that, even taking into account the subtleties of the preceding paragraph, the circuit $C^{f,p}_{n,e}$ has size polynomial in $n$ and $p$, and can be constructed in time polynomial in $n$ and $p$.

\section{Proof of Theorem~\ref{theorem:htr}}
\label{sec:proof}

Let $\mathcal{O}^{f,p}_{n,e}$ be constructed as described in Section~\ref{sec:oracle}. As is shown in~\cite{Boyer_1998}, a single run of Grover search with at most $O(2^{n/2})$ iterations is sufficient to determine, with success probability $O(1)$, whether the solution set is empty. By making use of the Chernoff-Hoeffding bound, one can show that the success probability can be boosted to $1-\delta'$ by repeating the Grover search $O(\log \delta'^{-1})$ times and taking a majority vote over successful runs. That a run is successful can be determined in polynomial time by checking whether the measurement outcome is indeed a bad rounding case. We denote this repeated Grover search as $\mathrm{QSearch}$ in Algorithm~\ref{alg:qhtr}. Each invocation of $\mathrm{QSearch}$ therefore makes at most $O(2^{n/2} \log \delta'^{-1})$ calls to an operator in $\mathcal{O}^{f,\bullet}_{n,e} = \{ \mathcal{O}^{f,p}_{n,e} \}_{p=n+1}^{p_{\max}}$. Since Algorithm~\ref{alg:qhtr} is a binary search, it makes at most $\lfloor \log p_{\max} \rfloor + 1$ iterations. For the algorithm to succeed with probability at least $1-\delta$, we must set a sufficiently small $\delta'$ for each iteration. Setting $\delta' \leq \delta / (\lfloor \log p_{\max} \rfloor + 1)$ suffices:
\begin{align}
    & \delta \geq \delta' (\lfloor \log p_{\max} \rfloor + 1) \ \Rightarrow \\ 
    & \Rightarrow 1-\delta \leq 1 - \delta' (\lfloor \log p_{\max} \rfloor + 1) \leq (1-\delta')^{\lfloor \log p_{\max} \rfloor + 1} \,,
\end{align}
where the last step is due to the Bernoulli inequality. Finally, if $p_{\max} = O(\mathrm{poly}(n))$ is an upper bound for $\mathrm{htr}_{f,I}(n)$, every operator in $\mathcal{O}^{f,\bullet}_{n,e}$ is implemented by a circuit of size $O(\mathrm{poly}(n))$ and the overall runtime is $\tilde O(2^{n/2} \log (1/\delta))$. \qed

\section{Oracle resource estimation for binary64}
\label{sec:resources}

\begin{table*}[t]
    \centering
    \begin{tabular}{c|ccccc}
        Circuit & $M(128)$ & $M(64)$ & $M(32)$ & $M(16)$ & $C^{\sin,106}_{53,e}$ \\
        \hline
        Schoolbook (gate count) & 65152 & 16192 & 4000 & 976 & $4.3 \times 10^6$ \\
        Karatsuba (gate count) & 31272 & 9912 & 3048 & 888 & $2.3 \times 10^6$ \\
        Parallel Karatsuba (depth) & 3112 & 1576 & 808 & 424 & $2.8 \times 10^4$ \\
    \end{tabular}
    \caption{Toffoli gate counts and parallelized Toffoli depths of $2^k$-bit multiplication circuits and the $C^{\sin,106}_{53,e}$ oracle.}
    \label{tab:multiplication}
\end{table*}

To translate this asymptotic advantage into a demonstration of practical quantum utility,
one promising target would be the rounding of $\sin(x)$ and $\cos(x)$ in the IEEE 754 standard
binary64 format. The bad cases for these functions for $x > 2^{11}$ were until this
spring~\cite{lefevre2026} out of reach for classical methods~\cite{brisebarre2025}.
Accordingly, we will consider the cost of implementing the required arithmetic circuits
for $C^{\sin,106}_{53,e}$ and the resulting cost of implementing Algorithm~\ref{alg:qhtr}.

The binary splitting algorithm~\cite{brent1976,Brent_Zimmermann_2010} can be used to calculate both
$\sin$ and $\cos$ at once. Its asymptotic complexity is less than the algebraic-geometric
mean algorithm, but below the $10^6$ bit range its smaller constants make it superior.
Achieving 106-bit precision requires partitioning the angle $x \mod 2\pi$ into eight
substrings $y_j$ with mantissas of length $1, 2, ..., 64, 128$. The binary splitting
algorithm can be applied separately (and in parallel) to calculate the sine and cosine of
each substring, and the results combined.

Evaluating $\sin y_j$ and $\cos y_j$ to the desired precision requires taking no more than
$\sqrt{256} = 16$ terms from the power series. Summing these terms via binary splitting's
divide-and-conquer method requires 45 multiplications and 15 additions, but only one
division. However, only three of these multiplications and one of these additions is at
the highest bit width. In total, the cost is $3 M(128) + 6 M(64) + 12 M(32) + 24 M(16)$
plus a negligible addition cost. Almost all of these operations can be implemented in
parallel, so the circuit depth required for one evaluation is only $M(128) + M(64) + M(32) + M(16)$.

Combining the results to obtain the final $\sin(x)$ and $\cos(x)$ requires an additional
$21 M(128)$ plus again a negligible addition cost. The total Toffoli cost is
$45 M(128) + 48 M(64) + 96 M(32) + 192 M(16)$; again, this can be parallelized heavily to
$8 M(128) + M(64) + M(32) + M(16)$.

With the schoolbook method of multiplication by shifted additions, $M(n) = 4n^2 - 3n$,
and we find the gate counts given in the first row of Tab.~\ref{tab:multiplication},
with the $C^{\sin,106}_{53,e}$ oracle subcircuit requiring a total of approximately
$4.3 \times 10^6$ Toffoli gates.

Especially for the larger multiplication circuits, the schoolbook method is
not the fastest available. Using reversible Karatsuba multiplication~\cite{parent2017},
we have $M(2^k) = 3 M(2^{k-1}) + 12 \cdot 2^k$. If we implement 8-bit multiplication
using the schoolbook method, $M(8) = 232$, yielding the gate counts in the second row of Tab.~\ref{tab:multiplication}. Thus the total
number of Toffoli gates required by $C^{\sin,106}_{53,e}$ is approximately
$2.3 \times 10^6$. However, by running each recursive Karatsuba multiplication subcircuit
in parallel, the circuit depth required is only $M(2^k) = M(2^{k-1}) + 12 \cdot 2^k$.
Using 8-bit schoolbook multiplication again yields the Toffoli depths in the third row of Tab.~\ref{tab:multiplication}. The depth could be reduced further by using schoolbook
multiplication only at the 4-bit level, but this would increase the Toffoli count.
The total Toffoli depth of the parallel $C^{\sin,106}_{53,e}$ is then approximately
$2.8 \times 10^4$.

Assuming an at least partially fault-tolerant quantum computer—as will likely be necessary
for gate counts even in the millions—the main limit on wallclock time will be the rate at
which syndromes can be extracted and decoded. With the very generous assumption of a
single-shot error correcting code~\cite{bombin2015} and factories directly distilling magic
Toffoli states from $T$-states~\cite{eastin2013}, the number of measurement cycles per Toffoli
gate can be as low as 1. On superconducting hardware, a measurement and repreparation can
require as little as hundreds of nanoseconds~\cite{spring2025}; thus the wallclock time per
oracle call could be as low as tens of milliseconds.

Algorithm~\ref{alg:qhtr} requires a number of oracle calls approximately equal to
$\frac{\pi}{4} 2^{n/2} \approx 7.45 \times 10^{7}$. To match the wallclock time of 4
hours per binade~\cite{lefevre2026}, the cost per oracle call would need to be reduced by two orders of magnitude, to
the hundreds of microseconds regime. On the other hand, searching all 1024 binades simultaneously would require only $\frac{\pi}{4} 2^{(n+10)/2} \approx 2.39 \times 10^{9}$ oracle calls; to match the total wallclock time of two weeks~\cite{lefevre2026}, the oracle would only need to run in hundreds of milliseconds, which would be possible for the architecture sketched above. This of course does not actually imply equal practical utility to the classical algorithm. First, the simultaneous search solves only the decision-problem version of hardness to round, which is equally difficult but far less useful. Second, the implementation described above would require a hypothetical quantum computer significantly larger and more advanced than any currently in existence, whereas the wallclock times stated for Lefèvre, Ly, and Zimmerman's algorithm are based on execution on extant classical computers~\cite{lefevre2026}. Still, it is valuable to consider what capabilities a quantum computer would require to make the table maker's quantum search a candidate for quantum advantage experiments.

\section{Discussion}
\label{sec:discussion}

Algorithm~\ref{alg:qhtr} with its $\tilde O(2^{n/2} \log (1/\delta))$ runtime achieves a modest asymptotic speedup compared to the best known classical algorithm or heuristic for computing the hardness to round periodic elementary functions in large binades, whose runtimes scale as $\tilde O(2^{4n/5})$ and $\tilde O(2^{(7-2\sqrt{10})n}) \approx \tilde O(2^{0.676n})$, respectively~\cite{Hanrot2007}. While there has been a long line of work on quantum arithmetic~\cite{wang2025comprehensive}, to our knowledge, this is the first work that demonstrates an asymptotic quantum speedup in a computer arithmetic task. What is more intriguing is that the speedup comes from standalone quantum search, which, as discussed in the introduction, is rather rare. Whether this speedup is fundamentally due to quantum-mechanical effects and therefore inaccessible to classical computers or simply an artifact of the existing classical algorithms is an open question, which we pose as a challenge to table makers.

Algorithm~\ref{alg:qhtr} has a number of shortcomings that can perhaps be overcome. First, it is exponential in $n$. The table maker's dilemma is related to factoring~\cite{Stehle2006}. Could Shor's~\cite{shor1994} or Regev's~\cite{regev2025} algorithms be adapted to this problem?

Second, the $\tilde O$ notation hides potentially large prefactors related to the complexity of the arithmetic circuits used. Any implementation would need to optimize the circuits considerably to fit them in the short coherence window of foreseeable quantum computers. On the other hand, for a given function, binade, and target precision, the hardness to round needs to be computed only once, thus amortizing the implementation cost. Given that, to date, we do not know the hardness to round for trigonometric functions in binades higher than $2^{11}$ for the binary64 format of the IEEE 754 standard~\cite{brisebarre2025}, it is conceivable that Algorithm~\ref{alg:qhtr} may be of some use eventually. In a practical scenario, Algorithm~\ref{alg:qhtr} can be sped up by focusing the search in the range $[n+1,2n]$ first, since the hardness to round is very likely in that range. Results in quantum arithmetic~\cite{wang2025comprehensive} could also be exploited to reduce the number of ancilla qubits required.

Third, while the algorithm can be generalized to determine the hardness to round over all binades simultaneously (instead of fixing the exponent, initialize a superposition of all possible exponent values), the arithmetic circuits for function evaluation will be more complex in this case, since different argument reductions are favorable in large and small binades~\cite{Brent_Zimmermann_2010} and hence the suitable logic will have to be built into the circuits. Still, the arithmetic circuits will remain poly-sized.

Finally, whether more efficient quantum algorithms for computing $\mathrm{htr}_{f,I}$ (and finding hard inputs) can be obtained by combining quantum search with suitable classical algorithms remains an open question.

\bibliography{refs}

@inproceedings{Amla2005,
author = {Amla, Nina and Du, Xiaoqun and Kuehlmann, Andreas and Kurshan, Robert P. and McMillan, Kenneth L.},
title = {An analysis of SAT-based model checking techniques in an industrial environment},
year = {2005},
isbn = {3540291059},
publisher = {Springer-Verlag},
address = {Berlin, Heidelberg},
url = {https://doi.org/10.1007/11560548_20},
doi = {10.1007/11560548_20},
booktitle = {Proceedings of the 13 IFIP WG 10.5 International Conference on Correct Hardware Design and Verification Methods},
pages = {254–268},
numpages = {15},
location = {Saarbr\"{u}cken, Germany},
series = {CHARME'05}
}

@book {Niven1956,
    AUTHOR = {Niven, Ivan},
     TITLE = {Irrational numbers},
    SERIES = {The Carus Mathematical Monographs},
    VOLUME = {No. 11},
 PUBLISHER = {Mathematical Association of America, ; distributed by John
              Wiley \& Sons, Inc., New York},
      YEAR = {1956},
     PAGES = {xii+164},
   MRCLASS = {10.1X},
  MRNUMBER = {80123},
MRREVIEWER = {I.\ A.\ Barnett},
}

@article{Lehmer1933,
 ISSN = {00029890, 19300972},
 URL = {http://www.jstor.org/stable/2301023},
 author = {D. H. Lehmer},
 journal = {The American Mathematical Monthly},
 number = {3},
 pages = {165--166},
 publisher = {Mathematical Association of America},
 title = {A Note on Trigonometric Algebraic Numbers},
 urldate = {2025-10-31},
 volume = {40},
 year = {1933}
}

@misc{kahan2004,
      title={A logarithm too clever by half}, 
      author={Kahan, W.},
      year={2004},
      url={https://people.eecs.berkeley.edu/~wkahan/LOG10HAF.TXT}, 
}

@misc{nesterenko2000approximationvaluesexponentialfunction,
      title={On the approximation of the values of exponential function and logarithm by algebraic numbers}, 
      author={Yu. Nesterenko and M. Waldschmidt},
      year={2000},
      eprint={math/0002047},
      archivePrefix={arXiv},
      primaryClass={math.NT},
      url={https://arxiv.org/abs/math/0002047}, 
}

@inbook{Nielsen_Chuang_2010,
    place={Cambridge},
    title={Quantum search algorithms},
    booktitle={Quantum Computation and Quantum Information: 10th Anniversary Edition},
    publisher={Cambridge University Press}, author={Nielsen, Michael A. and Chuang, Isaac L.},
    year={2010},
    pages={248–276}}

@article{Boyer_1998,
   title={Tight Bounds on Quantum Searching},
   volume={46},
   ISSN={1521-3978},
   url={http://dx.doi.org/10.1002/(SICI)1521-3978(199806)46:4/5<493::AID-PROP493>3.0.CO;2-P},
   DOI={10.1002/(sici)1521-3978(199806)46:4/5<493::aid-prop493>3.0.co;2-p},
   number={4–5},
   journal={Fortschritte der Physik},
   publisher={Wiley},
   author={Boyer, Michel and Brassard, Gilles and Høyer, Peter and Tapp, Alain},
   year={1998},
   month=jun, pages={493–505} }

@INPROCEEDINGS{shor1994,
  author={Shor, P.W.},
  booktitle={Proceedings 35th Annual Symposium on Foundations of Computer Science}, 
  title={Algorithms for quantum computation: discrete logarithms and factoring}, 
  year={1994},
  volume={},
  number={},
  pages={124-134},
  doi={10.1109/SFCS.1994.365700}}

@INPROCEEDINGS{Lefevre2005,
  author={Lefevre, V.},
  booktitle={17th IEEE Symposium on Computer Arithmetic (ARITH'05)}, 
  title={New Results on the Distance between a Segment and Z ² . Application to the Exact Rounding}, 
  year={2005},
  volume={},
  number={},
  pages={68-75},
  doi={10.1109/ARITH.2005.32}}

@ARTICLE{slz2005,
  author={Stehle, D. and Lefevre, V. and Zimmermann, P.},
  journal={IEEE Transactions on Computers}, 
  title={Searching worst cases of a one-variable function using lattice reduction}, 
  year={2005},
  volume={54},
  number={3},
  pages={340-346},
  doi={10.1109/TC.2005.55}}

@INPROCEEDINGS{Hanrot2007,
  author={Hanrot, Guillaume and Lefevre, Vincent and Stehle, Damien and Zimmermann, Paul},
  booktitle={18th IEEE Symposium on Computer Arithmetic (ARITH '07)}, 
  title={Worst Cases of a Periodic Function for Large Arguments}, 
  year={2007},
  volume={},
  number={},
  pages={133-140},
  doi={10.1109/ARITH.2007.37}}

@InProceedings{Coppersmith1996,
author="Coppersmith, Don",
editor="Maurer, Ueli",
title="Finding a Small Root of a Bivariate Integer Equation; Factoring with High Bits Known",
booktitle="Advances in Cryptology --- EUROCRYPT '96",
year="1996",
publisher="Springer Berlin Heidelberg",
address="Berlin, Heidelberg",
pages="178--189",
isbn="978-3-540-68339-1"
}

@InProceedings{Stehle2006,
author="Stehl{\'e}, Damien",
editor="Hess, Florian
and Pauli, Sebastian
and Pohst, Michael",
title="On the Randomness of Bits Generated by Sufficiently Smooth Functions",
booktitle="Algorithmic Number Theory",
year="2006",
publisher="Springer Berlin Heidelberg",
address="Berlin, Heidelberg",
pages="257--274",
isbn="978-3-540-36076-6"
}

@article{brisebarre2025,
author = {Brisebarre, Nicolas, PhD and Hanrot, Guillaume and Muller, Jean-Michel and Zimmermann, Paul},
title = {Correctly-Rounded Evaluation of a Function: Why, How, and at What Cost?},
year = {2025},
issue_date = {January 2026},
publisher = {Association for Computing Machinery},
address = {New York, NY, USA},
volume = {58},
number = {1},
issn = {0360-0300},
url = {https://doi.org/10.1145/3747840},
doi = {10.1145/3747840},
journal = {ACM Comput. Surv.},
month = sep,
articleno = {27},
numpages = {34}
}

@article{or1992patriot,
  author={{US General Accounting Office}},
  title={{Patriot Missile Defense: Software Problem Led to System Failure at Dhahran, Saudi Arabia}},
  journal={Report Number: GAO/IMTEC-92-26},
  year={1992}
}

@article{schmitt1991after,
  title={AFTER THE WAR; Army Is Blaming Patriot's Computer For Failure to Stop the Dhahran Scud.},
  author={Schmitt, Eric},
  journal={The New York Times},
  pages={NA--NA},
  year={1991},
  publisher={The New York Times Company}
}

@article{mccullough1999,
Author = {McCullough, B. D. and Vinod, H. D.},
Title = {The Numerical Reliability of Econometric Software},
Journal = {Journal of Economic Literature},
Volume = {37},
Number = {2},
Year = {1999},
Month = {June},
Pages = {633–665},
DOI = {10.1257/jel.37.2.633},
URL = {https://www.aeaweb.org/articles?id=10.1257/jel.37.2.633}}

@article{nievergelt2000rounding,
  title={Rounding Errors to Knock Your Stocks Off},
  author={Nievergelt, Yves},
  journal={Mathematics Magazine},
  volume={73},
  number={1},
  pages={47--48},
  year={2000},
  publisher={Taylor \& Francis}
}

@ARTICLE{Vizel2015,
  author={Vizel, Yakir and Weissenbacher, Georg and Malik, Sharad},
  journal={Proceedings of the IEEE}, 
  title={Boolean Satisfiability Solvers and Their Applications in Model Checking}, 
  year={2015},
  volume={103},
  number={11},
  pages={2021-2035},
  doi={10.1109/JPROC.2015.2455034}}

@Inbook{Bernstein2009,
author="Bernstein, Daniel J.",
title="Introduction to post-quantum cryptography",
bookTitle="Post-Quantum Cryptography",
year="2009",
publisher="Springer Berlin Heidelberg",
address="Berlin, Heidelberg",
pages="1--14",
isbn="978-3-540-88702-7",
doi="10.1007/978-3-540-88702-7_1",
url="https://doi.org/10.1007/978-3-540-88702-7_1"
}

@book{Arora_Barak_2009, place={Cambridge}, title={Computational Complexity: A Modern Approach}, publisher={Cambridge University Press}, author={Arora, Sanjeev and Barak, Boaz}, year={2009}}

@InProceedings{grassl2016,
author="Grassl, Markus
and Langenberg, Brandon
and Roetteler, Martin
and Steinwandt, Rainer",
editor="Takagi, Tsuyoshi",
title="Applying Grover's Algorithm to AES: Quantum Resource Estimates",
booktitle="Post-Quantum Cryptography",
year="2016",
publisher="Springer International Publishing",
address="Cham",
pages="29--43",
isbn="978-3-319-29360-8"
}

@inproceedings{grover1996,
author = {Grover, Lov K.},
title = {A fast quantum mechanical algorithm for database search},
year = {1996},
isbn = {0897917855},
publisher = {Association for Computing Machinery},
address = {New York, NY, USA},
url = {https://doi.org/10.1145/237814.237866},
doi = {10.1145/237814.237866},
booktitle = {Proceedings of the Twenty-Eighth Annual ACM Symposium on Theory of Computing},
pages = {212–219},
numpages = {8},
location = {Philadelphia, Pennsylvania, USA},
series = {STOC '96}
}

@article{bennett1989,
author = {Bennett, Charles H.},
title = {Time/Space Trade-Offs for Reversible Computation},
journal = {SIAM Journal on Computing},
volume = {18},
number = {4},
pages = {766-776},
year = {1989},
doi = {10.1137/0218053},
URL = {https://doi.org/10.1137/0218053},
}

@misc{Brassard_2002,
   title={Quantum amplitude amplification and estimation},
   ISSN={0271-4132},
   url={http://dx.doi.org/10.1090/conm/305/05215},
   DOI={10.1090/conm/305/05215},
   journal={Quantum Computation and Information},
   publisher={American Mathematical Society},
   author={Brassard, Gilles and Høyer, Peter and Mosca, Michele and Tapp, Alain},
   year={2002},
   pages={53–74} }

@article{levine1990,
author = {Levine, Robert Y. and Sherman, Alan T.},
title = {A Note on Bennett’s Time-Space Tradeoff for Reversible Computation},
journal = {SIAM Journal on Computing},
volume = {19},
number = {4},
pages = {673-677},
year = {1990},
doi = {10.1137/0219046},
URL = {https://doi.org/10.1137/0219046},
}

@ARTICLE{ieee754,
  author={Microprocessor Standards Committee},
  journal={IEEE Std 754-2019 (Revision of IEEE 754-2008)}, 
  title={{IEEE} Standard for Floating-Point Arithmetic}, 
  year={2019},
  volume={},
  number={},
  pages={1-84},
  doi={10.1109/IEEESTD.2019.8766229}}

@article{ambainis2004,
author = {Ambainis, A.},
title = {Quantum search algorithms},
year = {2004},
issue_date = {June 2004},
publisher = {Association for Computing Machinery},
address = {New York, NY, USA},
volume = {35},
number = {2},
issn = {0163-5700},
url = {https://doi.org/10.1145/992287.992296},
doi = {10.1145/992287.992296},
journal = {SIGACT News},
month = jun,
pages = {22–35},
numpages = {14}
}

@InProceedings{furer2008,
author="F{\"u}rer, Martin",
editor="Laber, Eduardo Sany
and Bornstein, Claudson
and Nogueira, Loana Tito
and Faria, Luerbio",
title="Solving NP-Complete Problems with Quantum Search",
booktitle="LATIN 2008: Theoretical Informatics",
year="2008",
publisher="Springer Berlin Heidelberg",
address="Berlin, Heidelberg",
pages="784--792",
isbn="978-3-540-78773-0"
}

@Inbook{muller2009solving,
author="Muller, Jean-Michel
and Brunie, Nicolas
and de Dinechin, Florent
and Jeannerod, Claude-Pierre
and Joldes, Mioara
and Lef{\`e}vre, Vincent
and Melquiond, Guillaume
and Revol, Nathalie
and Torres, Serge",
title="Evaluating Floating-Point Elementary Functions",
bookTitle="Handbook of Floating-Point Arithmetic",
year="2018",
publisher="Springer International Publishing",
address="Cham",
pages="375--433",
isbn="978-3-319-76526-6",
doi="10.1007/978-3-319-76526-6_10",
url="https://doi.org/10.1007/978-3-319-76526-6_10"
}

@inbook{Brent_Zimmermann_2010, place={Cambridge}, series={Cambridge Monographs on Applied and Computational Mathematics}, title={Elementary and special function evaluation}, booktitle={Modern Computer Arithmetic}, publisher={Cambridge University Press}, author={Brent, Richard P. and Zimmermann, Paul}, year={2010}, pages={125–184}, collection={Cambridge Monographs on Applied and Computational Mathematics}}

@article{regev2025,
author = {Regev, Oded},
title = {An Efficient Quantum Factoring Algorithm},
year = {2025},
issue_date = {February 2025},
publisher = {Association for Computing Machinery},
address = {New York, NY, USA},
volume = {72},
number = {1},
issn = {0004-5411},
url = {https://doi.org/10.1145/3708471},
doi = {10.1145/3708471},
journal = {J. ACM},
month = jan,
articleno = {10},
numpages = {13}
}

@article{wang2025comprehensive,
  title={A comprehensive study of quantum arithmetic circuits},
  author={Wang, Siyi and Li, Xiufan and Lee, Wei Jie Bryan and Deb, Suman and Lim, Eugene and Chattopadhyay, Anupam},
  journal={Philosophical Transactions A},
  volume={383},
  number={2288},
  pages={20230392},
  year={2025},
  publisher={The Royal Society}
}

@InProceedings{parent2017,
  author =	{Parent, Alex and Roetteler, Martin and Mosca, Michele},
  title =	{{Improved reversible and quantum circuits for Karatsuba-based integer multiplication}},
  booktitle =	{12th Conference on the Theory of Quantum Computation, Communication and Cryptography (TQC 2017)},
  pages =	{7:1--7:15},
  series =	{Leibniz International Proceedings in Informatics (LIPIcs)},
  ISBN =	{978-3-95977-034-7},
  ISSN =	{1868-8969},
  year =	{2018},
  volume =	{73},
  editor =	{Wilde, Mark M.},
  publisher =	{Schloss Dagstuhl -- Leibniz-Zentrum f{\"u}r Informatik},
  address =	{Dagstuhl, Germany},
  URL =		{https://drops.dagstuhl.de/entities/document/10.4230/LIPIcs.TQC.2017.7},
  URN =		{urn:nbn:de:0030-drops-85841},
  doi =		{10.4230/LIPIcs.TQC.2017.7}
}

@inproceedings{brent1976,
    author = {Brent, Richard P.},
    title = {The complexity of multiple-precision arithmetic},
    booktitle = {The Complexity of Computational Problem Solving},
    year = {1976},
    editor ={Anderssen, R. S. and Brent, R. P.},
    publisher = {University of Queensland Press},
    address = {Brisbane, Australia},
    pages = {126-165},
    url = {https://arxiv.org/abs/1004.3412}
}

@inproceedings{lefevre2026,
  TITLE = {{Computing hard-to-round cases of sin, cos, tan in double precision}},
  AUTHOR = {Lef{\`e}vre, Vincent and Ly, Tue and Zimmermann, Paul},
  URL = {https://inria.hal.science/hal-05593313},
  BOOKTITLE = {{ARITH 2026 - 33rd IEEE International Symposium on Computer Arithmetic}},
  ADDRESS = {Fulda, Germany},
  YEAR = {2026},
  MONTH = Jun,
  HAL_ID = {hal-05593313},
  HAL_VERSION = {v1},
}

@article{bombin2015,
  title = {Single-Shot Fault-Tolerant Quantum Error Correction},
  author = {Bomb\'{\i}n, H\'ector},
  journal = {Phys. Rev. X},
  volume = {5},
  issue = {3},
  pages = {031043},
  numpages = {26},
  year = {2015},
  month = {Sep},
  publisher = {American Physical Society},
  doi = {10.1103/PhysRevX.5.031043},
  url = {https://link.aps.org/doi/10.1103/PhysRevX.5.031043}
}

@article{eastin2013,
  title = {Distilling one-qubit magic states into Toffoli states},
  author = {Eastin, Bryan},
  journal = {Phys. Rev. A},
  volume = {87},
  issue = {3},
  pages = {032321},
  numpages = {7},
  year = {2013},
  month = {Mar},
  publisher = {American Physical Society},
  doi = {10.1103/PhysRevA.87.032321},
  url = {https://link.aps.org/doi/10.1103/PhysRevA.87.032321}
}

@article{spring2025,
  title = {Fast Multiplexed Superconducting-Qubit Readout with Intrinsic Purcell Filtering Using a Multiconductor Transmission Line},
  author = {Spring, Peter A. and Milanovic, Luka and Sunada, Yoshiki and Wang, Shiyu and van Loo, Arjan F. and Tamate, Shuhei and Nakamura, Yasunobu},
  journal = {PRX Quantum},
  volume = {6},
  issue = {2},
  pages = {020345},
  numpages = {23},
  year = {2025},
  month = {Jun},
  publisher = {American Physical Society},
  doi = {10.1103/PRXQuantum.6.020345},
  url = {https://link.aps.org/doi/10.1103/PRXQuantum.6.020345}
}
\end{document}